\documentclass{epsconf}
\usepackage{epsfig} % use this package to include EPS format figures
\usepackage{wrapfig}
\usepackage{amsmath}
\usepackage{amssymb}
\usepackage{amsfonts}
\usepackage{times}
\usepackage{subfigure}

\renewcommand{\phi}{\varphi}
\renewcommand{\epsilon}{\varepsilon}
\newcommand{\grad}{\nabla}
\newcommand{\Dd}{\ensuremath{\mathrm{d}}}

\renewcommand{\vec}[1]{\ensuremath{\boldsymbol{#1}}} % bold vectors, instead
\newcommand{\pp}[2]{\frac{\partial #1}{\partial #2}}
\newcommand{\dd}[2]{\frac{\Dd #1}{\Dd #2}}

\title{Impurity transport in ITG and TE mode dominated turbulence}
\author{\underline{A. Skyman}$^1$, H. Nordman$^1$, P. Strand$^1$, F. Jenko$^2$, F. Merz$^2$}
\institute{
  $^1$ Euratom-VR Association, Department of Radio and Space Science, Chalmers University
    of Technology, SE-412 96 G\"oteborg, Sweden.\\
  $^2$ Max-Planck-Institut f\"ur Plasmaphysik EURATOM-IPP, D-85748 Garching, Germany.
}

\begin{document}
\maketitle

\section{Introduction}
\label{sec:introduction}
\begin{wrapfigure}{r}{0.38\textwidth}
 \centering
 \vspace{-35pt}
 \includegraphics[width=\linewidth]{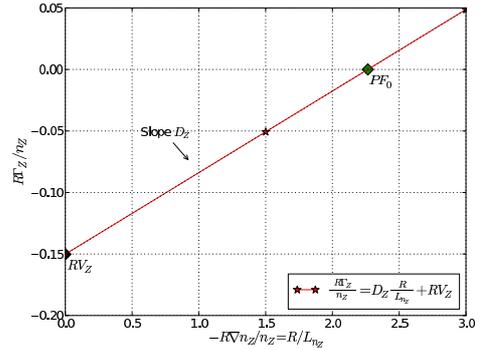}
 \caption[$Z$-scaling]{\footnotesize Illustration of $PF_0$ and the linearity
of $\Gamma_Z\left(\grad n_Z\right)$; ITG dominated quasilinear GENE result for $Ne$ with $k\rho = 0.3$}
 \label{fig:G}
 \vspace{-15pt}
\end{wrapfigure}
\noindent The transport properties of impurities is of high relevance for the
performance and optimisation of magnetic fusion devices. For instance, if
impurities from the plasma-facing surfaces accumulate in the core,
wall-impurities of relatively low density suffice to dilute the plasma and lead
to unacceptable energy losses in the form of radiation. 

In the present study, turbulent impurity transport in Deuterium tokamak plasmas, driven by
Ion Temperature Gradient (ITG) and Trapped Electron (TE) modes, has been investigated
using fluid and gyrokinetic models. The impurity diffusivity ($D_Z$) and convective
velocity ($V_Z$) are calculated, and from these the zero-flux peaking factor
($PF_0$) is derived. This quantity expresses the impurity density gradient at
which the convective and diffusive transport of impurities are exactly balanced.
The sign of $PF_0$ is of special interest, as it determines whether the
impurities are subject to an inward ($PF_0 > 0$) or outward ($PF_0 < 0$) pinch. 

Quasilinear results obtained from the GENE code \cite{Jenko2000, Merz2008} are
compared with two-fluid results \cite{Weiland2000} for both ITG and TE mode
dominated turbulence. Scalings of $PF_0$ with impurity
charge ($Z$) and various plasma parameters, such as magnetic shear ($\hat
s$), are studied. Of particular interest are conditions favouring an outward
convective impurity flux. 

\section{Theoretical background}
\noindent The transport of a trace impurity species can locally
be described by a \emph{diffusive} and a \emph{convective} part.
The former is characterized by the diffusion coefficient $D_Z$, the latter by a
convective velocity or ``pinch'' $V_Z$, see
equation~\eqref{eq:transport}~\cite{Angioni2006}.
From these, the \emph{zero flux peaking factor} is defined as
$PF_0=\frac{-R\,V_Z}{D_Z}|_{\Gamma = 0}$, see figure~\ref{fig:G}.
$PF_0$ is important in reactor design,  as it quantifies the
balance of convective and diffusive transport. This can be seen from equation
\eqref{eq:transport}, where $\Gamma_Z$ is the impurity flux, $n_Z$ the density
of the impurity species and $R$ the major radius of the
tokamak.  For the domain studied -- a narrow flux tube --
the gradient of the impurity density is constant: $\grad n_Z/n_Z=1/L_{n_Z}$. Setting
$\Gamma_Z = 0$ in equation~\eqref{eq:transport} yields the interpretation of 
$PF_0$ as the gradient of zero impurity flux.

\begin{equation}
  \label{eq:transport}
  \Gamma_Z = -D_Z\grad n_Z + n_Z V_Z \Leftrightarrow \frac{R\Gamma}{n_Z} =
-D_Z\frac{R}{L_{n_Z}} + RV_Z
\end{equation}

\section[Fluid model]{Fluid model}
\noindent Though the main results presented in this study have been obtained using quasilinear gyrokinetic simulations, their physical meaning is interpreted by comparing with the Weiland multi-fluid model~\cite{Weiland2000}.  The fluid equations for each included species ($j = i,\, te,\, Z$, representing Deuterium ions, trapped electrons, and trace impurities) are:

\vspace{-1\bigskipamount}
\begin{eqnarray}
 \pp{n_j}{t} + \grad \cdot \left( n_j \vec{v}_j \right)=0
 \label{eq:continuity} \\
 m_{i,Z} n_{i,Z} \pp{v_{||i,Z}}{t} + \grad_{||} \left( n_{i,Z} T_{i,Z} \right) +
n_{i,Z} e \grad_{||} \phi = 0 \label{eq:parallelmotion} \\
 \frac{3}{2} n_j \dd{T_j}{t} + n_j T_j \grad \cdot \vec{v}_j + \grad \cdot
\vec{q}_j = 0 \label{eq:temperature}
\end{eqnarray}

\noindent Here $\vec{q}_j$ is the diamagnetic heat flux, and $\vec{v_j}$ is the sum of the $\vec{E}\times\vec{B}$, diamagnetic drift, polarization drift, and stress-tensor drift velocities. To solve the equations, it is assumed that $\vec{q}_j$ is the only heat flux for all species, that passing electrons are adiabatic, and that quasineutraility (equation~\eqref{eq:qneutrality}) holds. Going to the trace limit for the impurities, i.e. letting $Zf_Z\rightarrow 0$ in equation~\eqref{eq:qneutrality}, an eigenvalue equation for ITG and TE modes is obtained. The impurity particle flux in equation~\eqref{eq:transport} is then obtained from $\Gamma_{nj}=\langle\delta n_j\vec{v}_{\vec{E}\times\vec{B}}\rangle$, where the averaging is performed over all unstable modes for a fixed length scale $k\rho$ of the turbulence.

\vspace{-1\bigskipamount}
\begin{equation}
 \label{eq:qneutrality}
 \frac{\delta n_e}{n_e} = \left(1 - Z f_Z\right) \frac{\delta n_i}{n_i} + Z
 f_Z \frac{\delta n_Z}{n_Z},\enspace f_Z = \frac{n_Z}{n_e}
\end{equation}

\section[Simulations]{Quasilinear gyrokinetic simulations}
\noindent GENE is a parallel gyrokinetic code employing a fixed grid in five dimensional phase space
and a flux-tube geometry~\cite{Jenko2000}. The simulations were performed on the \emph{HPC-FF} cluster\footnote{HPC-FF (\emph{High Performance Computing For Fusion}) is an EFDA funded computer situated at Forschungs\-zentrum J\"ulich. Germany, dedicated to fusion research} with GENE running in eigenvalue mode. Growth rates and impurity fluxes were thus computed for ITG and TE mode dominated cases, for which a number of parameters were varied and trends observed. The main parameters used are presented in table~\ref{tab:parameters}.

\begin{table}[ht]
 \centering
 \scriptsize
 \caption[Parameters]{\small Parameters used in all simulations}
 \label{tab:parameters}
 \begin{tabular}{l||r|r}\hline
 & ITG: & TEM: \\ \hline\hline
 $T_D/T_e$:             & $1.0$   & $1.0$\\
 $\hat{s}$:             & $0.8$   & $0.8$\\
 $q_0$:                 & $1.4$   & $1.4$\\
 $\epsilon$:            & $0.14$  & $0.14$\\
% $n_e$, $n_D+n_Z$: &$1.0$& $1.0$& $10^{19}$\unit{m^{-3}}\\
% $R$:           & $1.0$   & $1.0$&\unit{m}\\
 $R/L_{T_D},R/L_{T_Z}$: & $7.0$   & $3.0$\\
 $R/L_{T_e}$:           & $3.0$   & $7.0$\\ \hline
 % $\left(l_x,\, l_y\right)$:  & $\left(2\pi,\, 2\pi/k\rho\right)\cdot\rho$ & $\left(2\pi,\, 2\pi/k\rho\right)\cdot\rho$ \\
 $N_x\times N_{ky}\times N_z$: & $5\times1\times24$ & $4\times1\times24$\\
 $N_{v_{||}}\times N_\mu$:     & $64\times12$ & $64\times12$
 \end{tabular}
\end{table}

\section[Results]{Results}
\begin{figure}[ht]
 \centering
 \vspace{\smallskipamount}
 \subfigure[ITG mode dominated case]{\includegraphics[width=0.48\linewidth]{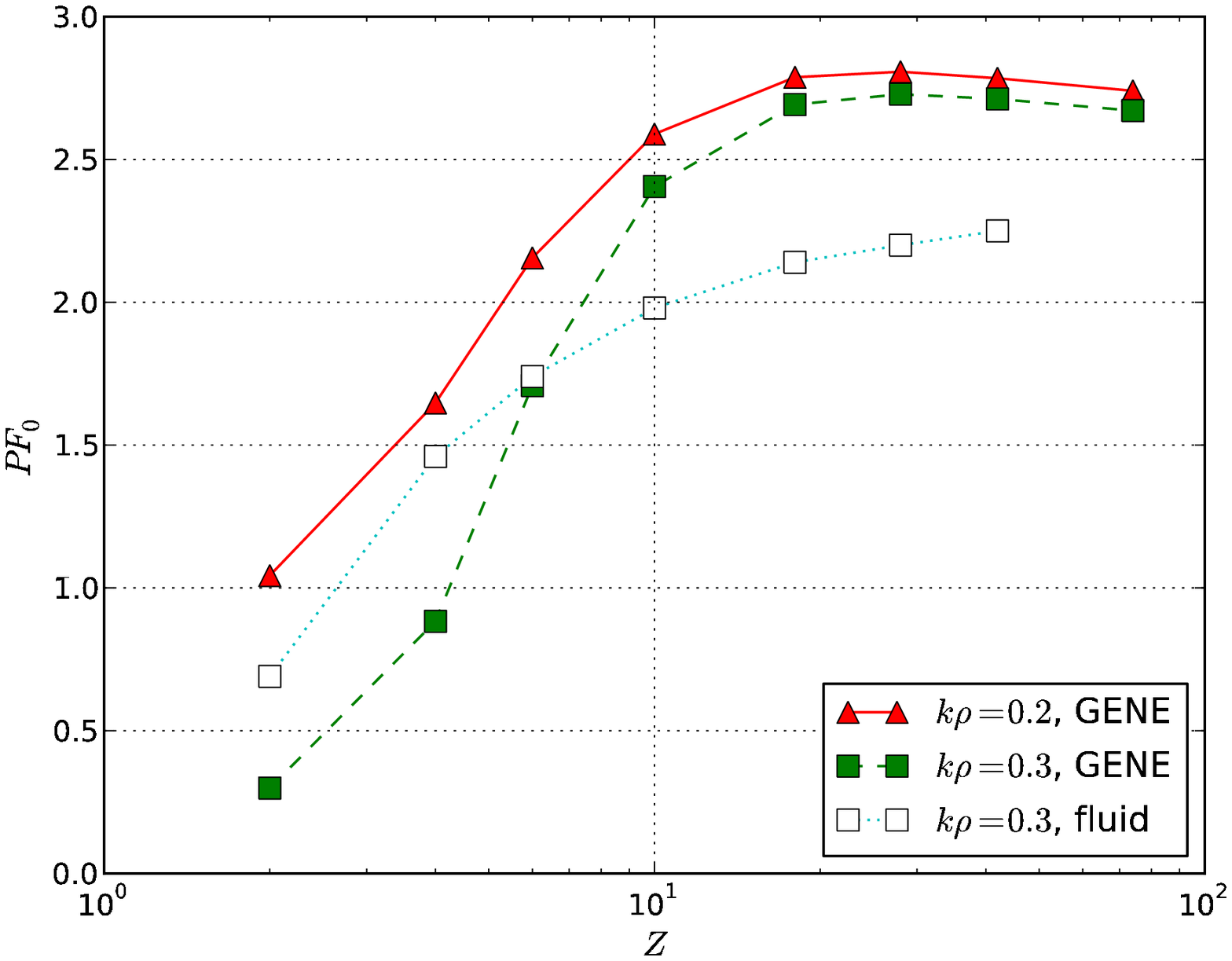}
\label{fig:Z_ITG}}
 ~
 \subfigure[TEM dominated case]{\includegraphics[width=0.48\linewidth]{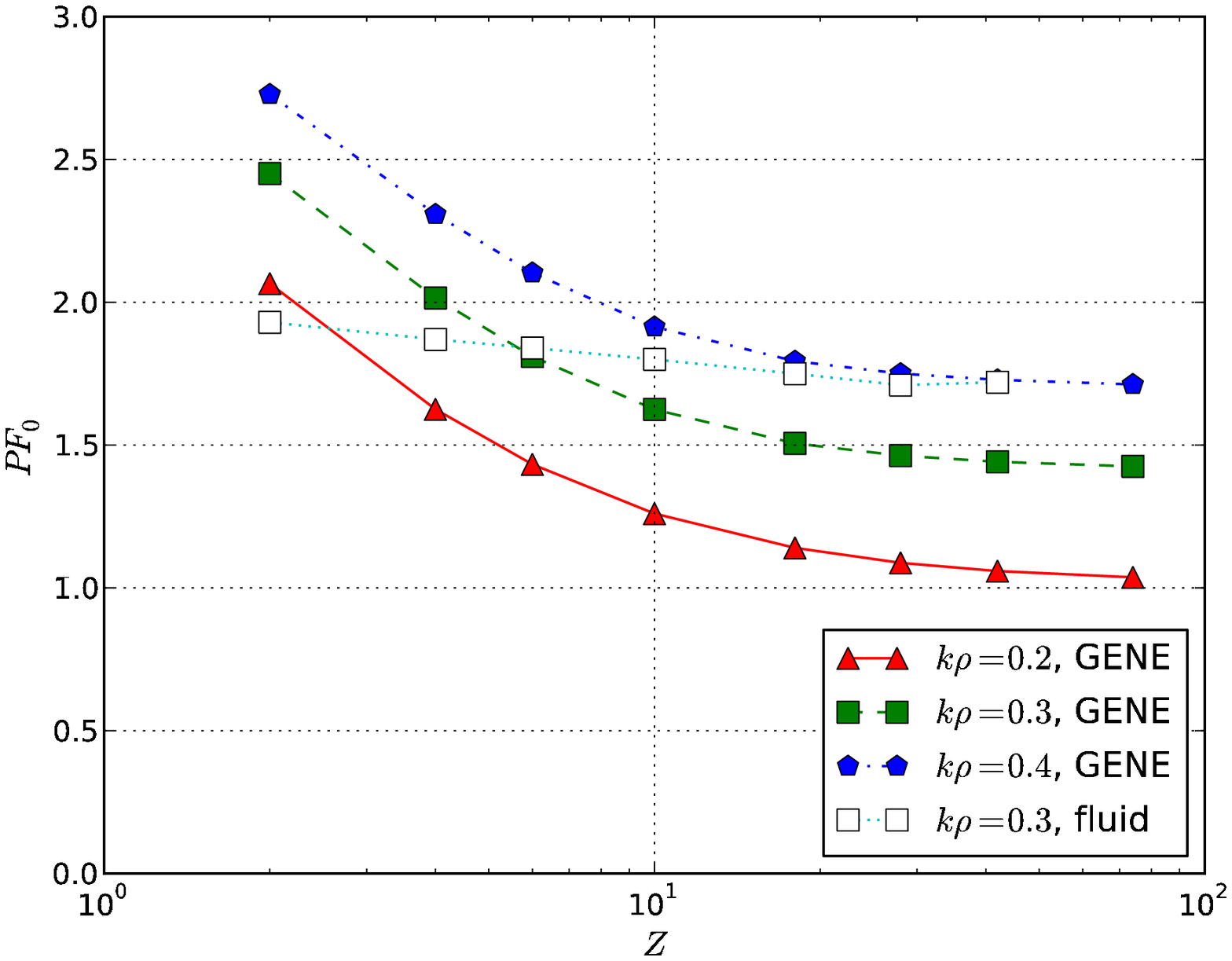}
\label{fig:Z_TEM}}
 \caption[$Z$-scaling]{\small Scalings of $PF_0$ with impurity charge $Z$; quasilinear GENE and fluid results}
 \label{fig:Z}
\end{figure}

\paragraph{Impurity charge $Z$:} The main results obtained are the scalings of the peaking factor with the charge of the impurity species. These are presented in figures~\ref{fig:Z_ITG} and~\ref{fig:Z_TEM}, showing ITG and TE mode dominated turbulence respectively.

\noindent The difference between figure~\ref{fig:Z_ITG} and \ref{fig:Z_TEM} can be understood from the properties of the convective velocity in~\eqref{eq:transport}. $V_Z$ contains a thermodiffusive term $V_{T_Z}\sim \frac{1}{Z}\frac{R}{L_{T_Z}}$ and a parallel impurity compression term $V_{p_Z}\sim\frac{Z}{A_Z}k_{||}^2\sim\frac{Z}{A_Zq^2}$. The former is generally outward ($V_{T_Z}>0$) for ITG and inward ($V_{T_Z}<0$) for TE mode dominated transport, whereas for the latter the opposite is generally the case. 
%For low $Z$ the thermal pinch dominates, leading to a lower $PF_0$ for ITG and a higher for TEM, but for higher $Z$ the parallell compression is more important, leading to a higher $PF_0$ for ITG and a lower for $TEM$ turbulence. 

\begin{figure}[t]
 \centering
 %\vspace{\smallskipamount}
 \subfigure[ITG mode dominated case]{\includegraphics[width=0.48\linewidth]{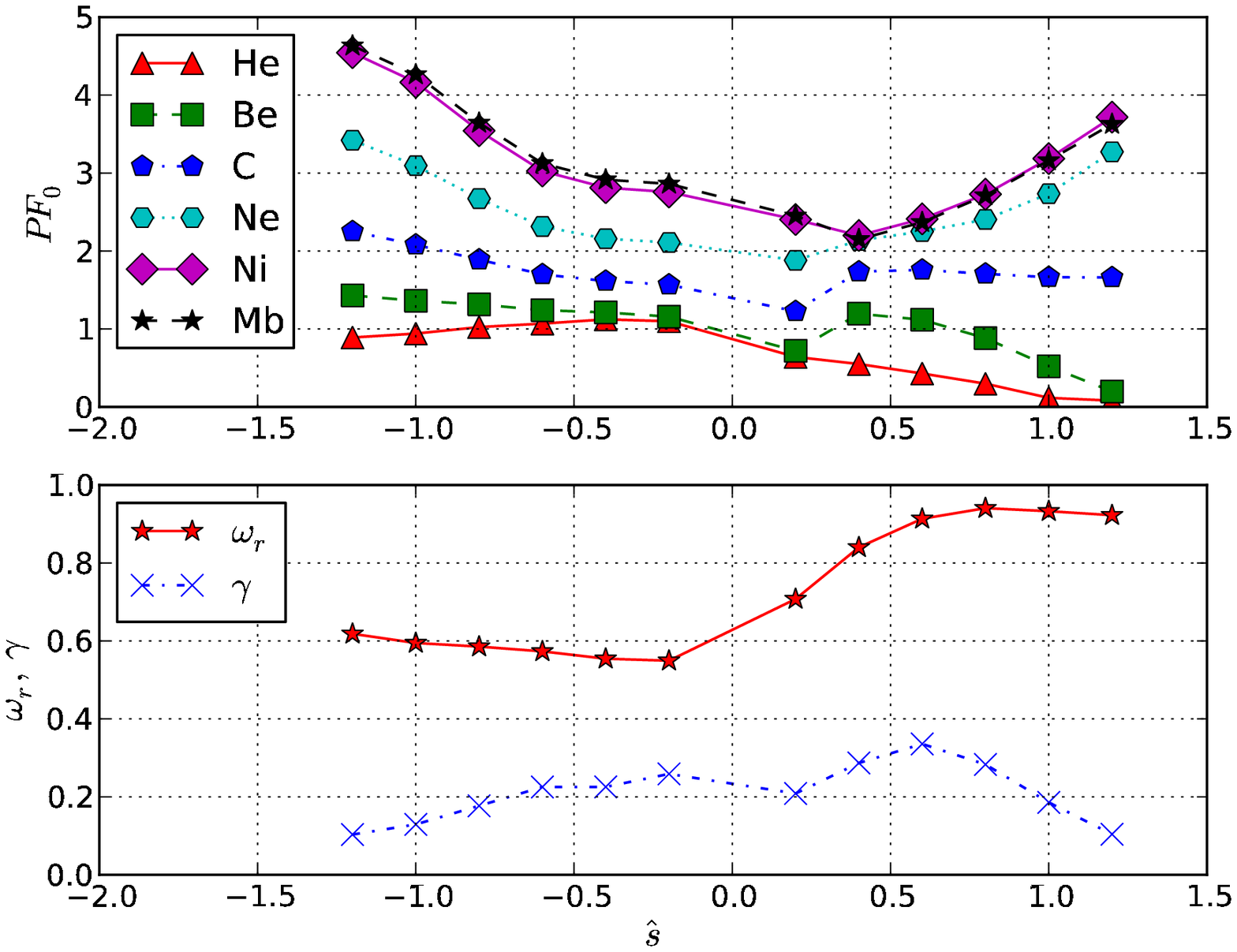}
\label{fig:s_ITG}}
 ~
 \subfigure[TEM dominated case]{\includegraphics[width=0.48\linewidth]{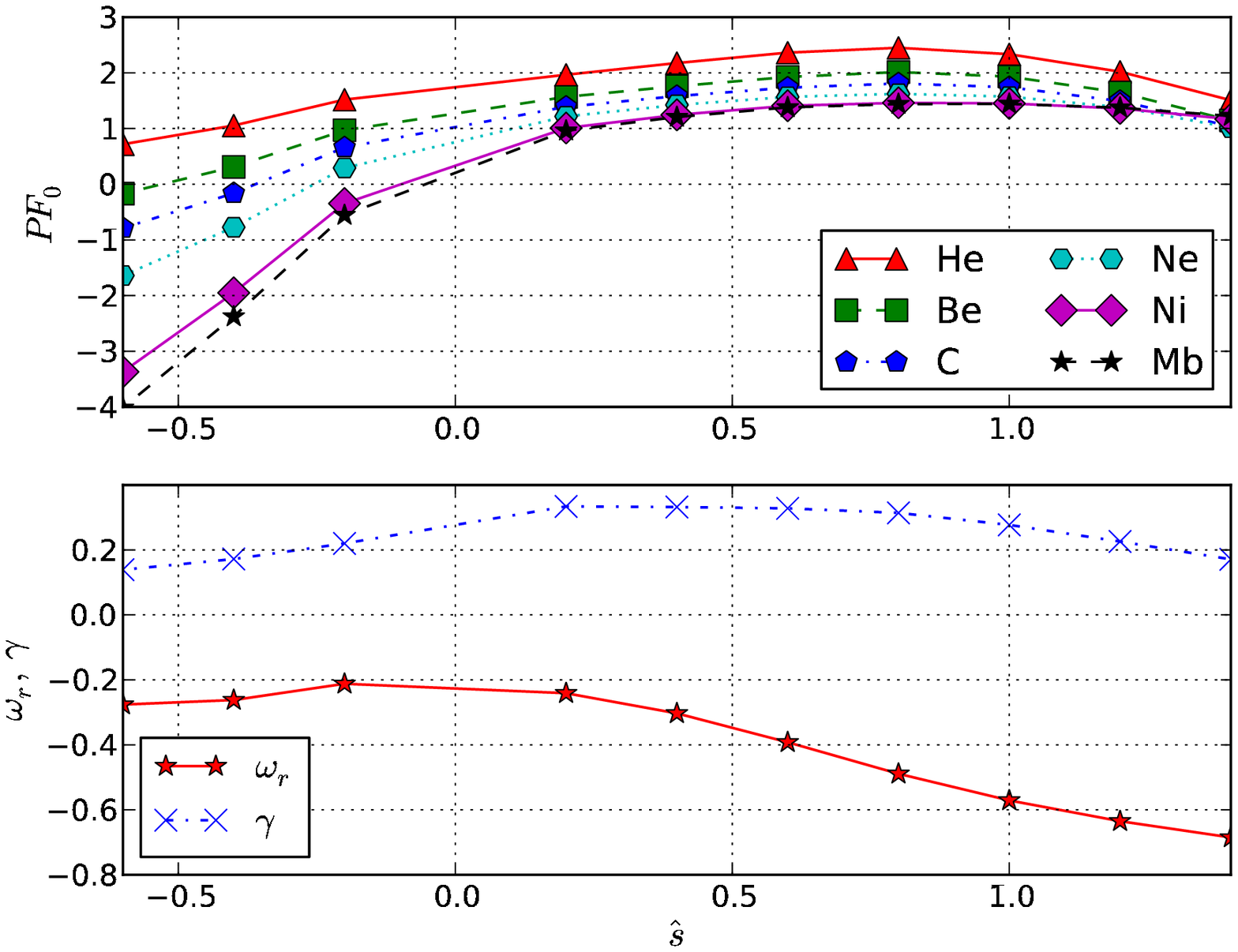}
\label{fig:s_TEM}}
 \caption[$\hat{s}$-scaling]{\small Scalings of $PF_0$ with magnetic shearing $\hat{s}$; quasilinear GENE results with $k\rho=0.3$}
 \label{fig:s}
\end{figure}

\vspace{-1\bigskipamount}
\paragraph{Magnetic shear $\hat{s}$:} The effect of magnetic shear on the peaking factor is shown in figures~\ref{fig:s_ITG} and~\ref{fig:s_TEM}. It is worth noting that a flux reversal, i.e. a change of sign in $PF_0$, owing to a change in sign of $V_Z$, occurs for negative $\hat{s}$ for $Z\gtrsim6$ in the TE mode dominated case, indicating a net outward transport of the heavier elements. Similar trends are not seen in fluid simulations, and this warrants further investigation.

\paragraph{Other parameters:} Scans of the dependence $PF_0$ on other parameters, such as $k\rho$ and $L_T$, have also been carried out. The results are similar to those reported in \cite{Dannert2005}, \cite{Nordman2008} and \cite{Fulop2009} respectively. In most cases, only a weak dependence of $PF_0$ is observed. 

% \begin{figure}[ht]
%  \centering
%  \includegraphics[width=0.618\linewidth]{krho_TEM.eps}
%  \caption[$k\rho$-scaling]{\small Scaling of $PF_0$ with $k\rho$ for TEM
% dominated cases; quasilinear GENE results}
%  \label{fig:krho_TEM}
% \end{figure}

% \begin{figure}[ht]
%  \centering
%  \includegraphics[width=0.618\linewidth]{omt_TEM.eps}
%  \caption[$L_{T_e}^{-1}$-scaling]{\small Scaling of $PF_0$ with $L_{T_e}^{-1}$
% for TEM dominated case; quasilinear GENE results}
%  \label{fig:omt_TEM}
% \end{figure}

\section[Conclusions]{Conclusions and outlook}
\noindent Quasilinear GENE simulations and fluid results show that peaking factor increases with impurity charge $Z$ for ITG mode dominated transport, whereas the opposite holds for TE mode dominated transport. In both cases $PF_0$ saturates for high $Z$. 

For magnetic shear, a flux reversal is observed for negative magnetic shear in the TEM dominated case. This is not seen in fluid simulations, and will be a focus of future studies.

For other parameters investigated, weak scalings for $PF_0$ are observed, in agreement with previous work.

\scriptsize
\bibliography{fusion.bib}

\begin{thebibliography}{1}

\bibitem{Jenko2000}
F.~Jenko, W.~Dorland, M.~Kotschenreuther, and B.~N. Rogers.
\newblock Electron temperature gradient driven turbulence.
\newblock {\em Physics of Plasmas}, 7(5):1904--10, May 2000.

\bibitem{Merz2008}
F.~Merz.
\newblock {\em Gyrokinetic Simulation of Multimode Plasma Turbulence}.
\newblock Monography, Westf\"alischen Wilhelms-Universit\"at M\"unster, 2008.

\bibitem{Weiland2000}
J.~Weiland.
\newblock {\em Collective Modes in Inhomogeneous Plasma}.
\newblock Institute of Physics Publishing, London, UK, 2000.

\bibitem{Angioni2006}
C.~Angioni and A.~G. Peeters.
\newblock Direction of impurity pinch and auxiliary heating in tokamak plasmas.
\newblock {\em PRL}, 96:095003--1--4, 2006.

\bibitem{Dannert2005}
T.~Dannert.
\newblock {\em Gyrokinetische {S}imulation von {P}lasmaturbulenz mit gefangenen
  {T}eilchen und elektromagnetischen {E}ffekten}.
\newblock Monography, Technischen {U}niversit\"at {M}\"unchen, January 2005.

\bibitem{Nordman2008}
H.~Nordman, R.~Singh, and T.~F\"ul\"op et~al.
\newblock Influence of the radio frequency ponderomotive force on anomalous
  impurity transport in tokamaks.
\newblock {\em PoP}, 15:042316--1--5, 2007.

\bibitem{Fulop2009}
T.~F\"ul\"op and H.~Nordman.
\newblock Turbulent and neoclassical impurity transport in tokamak plasmas.
\newblock {\em PoP}, 16:032306--1--8, 2009.

\end{thebibliography}
\bibliographystyle{unsrt}
\footnotesize
This work benefited from an allocation on the EFDA HPC-FF computer.

% 
% \begin{thebibliography}{99}
% \bibitem{GENE}
% A. First, B.C. Second and D. Third, Journal of interesting papers {\bf 10}, 10 (2004)
% \end{thebibliography}
% 
\end{document}